\documentclass[aps,prl,twocolumn,amssymb,nofootinbib]{revtex4}
\usepackage{graphicx}
\usepackage{amsmath}
\usepackage{amssymb}
\usepackage{bm}

\usepackage{epsfig}
\usepackage{bbm}
\usepackage{color}
\usepackage{ulem}

\definecolor{MyDarkBlue}{rgb}{0.15,0.15,0.45}
\usepackage[linktocpage=true]{hyperref}
\hypersetup{
colorlinks=true,
citecolor=MyDarkBlue,
linkcolor=MyDarkBlue,
urlcolor=MyDarkBlue,
pdfauthor={Melinda Andrews, Garrett Goon, Kurt Hinterbichler, James Stokes and Mark Trodden},
pdftitle={Massive gravity coupled to DBI Galileons is ghost free},
pdfsubject={hep-th}
}

\usepackage{enumitem}
\usepackage{graphicx}
\usepackage{subfigure}
\usepackage{amsmath}
\usepackage{bbm}
\usepackage{color}

\newcommand{\be}{\begin{equation}}
\newcommand{\ee}{\end{equation}}
\newcommand{\bea}{\begin{align}}
\newcommand{\eea}{\end{align}}

\newcommand{\rd}{{\rm d}}
\newcommand{\bE}{{E}}
\newcommand{\bF}{\bar{{E}}}

\begin{document}

\title{Massive Gravity Coupled to Galileons is Ghost Free} 
\author{Melinda Andrews$^{1}$\footnote{mgildner@sas.upenn.edu}}
\author{Garrett Goon$^{1}$\footnote{ggoon@sas.upenn.edu}}
\author{Kurt Hinterbichler$^{2}$\footnote{khinterbichler@perimeterinstitute.ca}}
\author{James Stokes$^{1}$\footnote{stokesj@sas.upenn.edu}}
\author{Mark Trodden$^{1}$\footnote{trodden@physics.upenn.edu}}

\affiliation{$^{1}$Center for Particle Cosmology, Department of Physics and Astronomy, University of Pennsylvania,
Philadelphia, Pennsylvania 19104, USA. \\
$^{2}$Perimeter Institute for Theoretical Physics, 31 Caroline St. N., Waterloo, ON, N2L 2Y5, Canada.}
\date{\today}

\begin{abstract}
It is possible to couple Dirac-Born-Infeld (DBI) scalars possessing generalized Galilean internal shift symmetries (Galileons) to nonlinear massive gravity in four dimensions, in such a manner that the interactions maintain the Galilean symmetry.  Such a construction is of interest because it is not possible to couple such fields to massless General Relativity in the same way.  We show that this theory has the primary constraint necessary to eliminate the Boulware-Deser ghost, thus preserving the attractive properties of both the Galileons and ghost-free massive gravity.
\end{abstract}

\maketitle

{\bf Introduction and Outline:}
It has long been known that the Fierz-Pauli action \cite{Fierz:1939ix} provides a consistent description of the linear fluctuations of a massive graviton in flat spacetime.  Nonlinear theories of massive gravity tend to suffer from an instability known as the Boulware-Deser (BD) ghost \cite{Boulware:1973my}.
A general Lagrangian for nonlinear massive gravity can be formulated by introducing a non-dynamical reference metric $\bar g_{\mu\nu}$ (e.g. the Minkowski one, $\bar g_{\mu\nu}=\eta_{\mu\nu}$) and constructing a potential of the form $V(g^{\mu\sigma}\bar g_{\sigma\nu})$. The potential explicitly breaks diffeomorphism invariance and it is expected that the theory generally propagates 12 phase-space degrees of freedom, rather than the 10 necessary to describe a massive graviton.  The extra degree of freedom is the BD ghost.  

The problem of finding a ghost-free nonlinear theory was only recently solved by de Rham, Gabadaze and Tolley (dRGT) \cite{deRham:2010ik,deRham:2010kj}.  dRGT theory is a 3-parameter family of potentials whose special structure ensures that there is a dynamical constraint which removes the ghost degree of freedom. This has been demonstrated by explicitly counting degrees of freedom in the Hamiltonian formalism \cite{Hassan:2011hr,Hassan:2011ea}, and through other methods \cite{deRham:2011qq,deRham:2011rn,Mirbabayi:2011aa,Hinterbichler:2012cn}. 

The diffeomorphism invariance broken by the mass term can be restored through the St\"uckelberg method \cite{ArkaniHamed:2002sp}, which involves introducing four auxiliary scalars $\phi^{\cal A}(x)$ through the replacement $\bar g_{\mu\nu} \rightarrow \partial_\mu \phi^{\cal A} \partial_\nu \phi^{\cal B} \eta_{\cal AB}.$  The St\"uckelberg fields are pure gauge and the original theory is recovered by choosing unitary gauge $\phi^{\cal A}=\delta^{\cal A}_\mu x^\mu$.
In this formulation, the scalars can be regarded as the embedding mapping of a sigma model $\Sigma \to \mathcal{M}$, where both $\Sigma$ and $ \mathcal{M} $ are four dimensional Minkowski space.  There is an internal Poincar\'{e} symmetry corresponding to the isometries of the target space.  From this point of view, the dynamical metric $g_{\mu\nu}(x)$ is a worldvolume metric living on $\Sigma$. 

The target space may be higher dimensional, and need not be flat -- we may generalize the sigma model to map to an arbitrary target space of dimension $D\geq 4$ with coordinates $\phi^{\cal A}$ (so that now ${\cal A},{\cal B},\cdots$ run over $D$ values) and a fixed target space metric $G_{{\cal A}{\cal B}}(\phi)$, 
\be \label{Inducedmet} \bar g_{\mu\nu} = \partial_\mu \phi^{\cal A} \partial_\nu \phi^{\cal B} G_{{\cal A}{\cal B}}(\phi).\ee  
There are now $(\dim \mathcal{M}) - 4$ embedding fields which cannot be gauged away and these become physical Dirac-Born-Infeld (DBI) scalars coupled to the physical metric through the dRGT potential \cite{Gabadadze:2012tr}. Apart from the dRGT terms, curvature invariants constructed solely from $\bar g_{\mu\nu}$ and extrinsic curvatures of the embedding can be included in the action.  The leading term in the derivative expansion is the DBI action $\sim\int {\rm d}^4 x\sqrt{-\bar g}$, and higher Lovelock invariants give Galileons \cite{deRham:2010eu,Hinterbichler:2010xn}.
The theory will possess a Galileon-like internal symmetry for every isometry of $G_{{\cal A}{\cal B}}$, and the resulting Galileons will be the generalized curved space Galileons discussed in \cite{Goon:2011qf,Goon:2011uw,Burrage:2011bt,toappear}.

Apart from generalizing dRGT, the construction of \cite{Gabadadze:2012tr} is of interest because it provides a method of coupling the Galileons to (massive) gravity while preserving the Galileon invariance.  When coupling to ordinary massless gravity, non-minimal couplings must be added to ensure second-order equations of motion, and the Galileon symmetry is broken \cite{Deffayet:2009wt,Deffayet:2009mn}.  In the present construction, there is no such problem, suggesting that the Galileons more naturally couple to a massive graviton.

In \cite{Gabadadze:2012tr}, it was shown that the theory is ghost-free, for a flat target space metric, in the decoupling limit, and for a certain simplifying choice of parameters.  In this letter, using methods similar to those of \cite{Hinterbichler:2012cn}, we demonstrate that the full theory, for any target space metric $G_{{\cal A}{\cal B}}$, to all orders beyond the decoupling limit, and for all choices of parameters, has the primary constraint necessary to eliminate the Boulware-Deser ghost.

{\bf The model:} The dynamical variables are the physical metric $g_{\mu\nu}$ and the $D$ scalars $\phi^{\cal A}$, which appear through the induced metric \eqref{Inducedmet}.  The action is
\begin{align}\label{actionm}
S&\!=\! S_{\rm EH}[g]+\!S_{\rm mix}[g,\bar g]+\!S_{\rm Galileon}[\bar g] \ .
\end{align}
Here $S_{\rm EH}[g]$ is the Einstein-Hilbert action for $g_{\mu\nu}$, with a possible cosmological constant $\Lambda$,
\be S_{\rm EH}[g]=\frac{M_{P}^{2}}{2}\!\int\!\rd^{4}x\, \sqrt{-g}\left(R[g]-2\Lambda\right).\ee
The action mixing the two metrics is
\begin{align}\label{mixaction}
S_{\rm mix}[g,\bar g]&=-{M_{P}^{2}m^{2}\over 8}\sum_{n=1}^{3}\beta_{n}S_n\left(\sqrt{g^{-1}\bar g}\right) \ ,
\end{align}
where $\sqrt{g^{-1}\bar g}$ is the matrix square root of the matrix $g^{\mu\sigma}\bar g_{\sigma \nu}$,
and $S_n(M)$ of a matrix $M$ are the symmetric polynomials\footnote{Our anti-symmetrization weight is $[\mu_1\ldots\mu_n]={1\over n!}(\mu_1\cdots\mu_n+\cdots)$.  See appendix A of ~\cite{Hinterbichler:2012cn} for more details on the symmetric polynomials.} $
S_n(M) = 
M^{[ \mu_1}_{\ \ \mu_1}\cdots M^{ \mu_n]}_{\ \ \mu_n}\,$.  The $\beta_n$ are three free parameters (one combination of which is redundant with the mass $m$).
$S_{\rm Galileon}[\bar g]$ stands for any Lagrangian constructed from diffeomorphism invariants of $\bar g$ (and extrinsic curvatures of the embedding) whose equations of motion remain second order in time derivatives.  The possible terms in $S_{\rm Galileon}[\bar g]$ are the Lovelock invariants and their boundary terms (see \cite{deRham:2010eu} and Sec. IV.B of \cite{Hinterbichler:2010xn} for a discussion).  The structure of the dRGT-DBI coupled system \eqref{actionm} is nearly identical to that of ghost-free bi-gravity \cite{Hassan:2011zd}, the difference  being that one of the two metrics is induced from a target space, and so it fundamentally depends on the embedding scalars.

Following \cite{Hinterbichler:2012cn}, we will find it convenient to write the theory in vierbein form\footnote{See also \cite{Deffayet:2012nr,Deser:2012qx} for covariant methods of degree of freedom counting in the vierbein formulation of massive gravity.}.  We write the physical metric and induced metric in terms of vierbeins $\bE^A=E_\mu^{\ A} \rd{ x}^\mu$, $\bF^A={\bar E}_\mu^{\ A} \rd{ x}^\mu$, 
\be g_{\mu\nu}=E_\mu^{\ A} E_\nu^{\ B}\eta_{AB}, \ \ \ \bar g_{\mu\nu}={\bar E}_\mu^{\ A} {\bar E}_\nu^{\ B}\eta_{AB},\ee
where $\eta_{AB}$ is the 4 dimensional Minkowski metric.
For the induced metric $\bar g_{\mu\nu}$, we write the vierbein in an upper triangular form 
 \be {\bar E}_{\mu}{}^{B}=\begin{pmatrix}
 {\bar N}& {\bar N}^{i}\bar e_{i}{}^{a}\\ 0 & \bar e_{i}{}^{a} \label{f:vierbeins}
\end{pmatrix} \ .
\ee
Here $\bar N$ and $\bar N^i$ are ADM lapse and shift variables, and $\bar e_{i}{}^{a}$ is an upper triangular spatial dreibein for the spatial part of the induced metric and $\bar e^{i}{}_{a}$ its inverse transpose (in what follows $i,j,\ldots $ are spatial coordinate indices raised and lowered with the spatial metric $\bar g_{ij}$, and $a,b,\ldots$ are spatial Lorentz indices raised and lowered with $\delta_{ab}$).  These are obtained in terms of $\phi^{\cal A}$ by solving
\begin{align}
{\bar g}_{00}=\dot{\phi}^{\cal A}\dot{\phi}^{\cal B}G_{{\cal A}{\cal B}}(\phi)&=- {\bar N}^{2}+ {\bar N}^{i} {\bar N}_{i} \nonumber\\
{\bar g}_{0i}=\dot{\phi}^{\cal A}\partial_{i}\phi^{\cal B}G_{{\cal A}{\cal B}}(\phi)&= {\bar N}_{i} \nonumber\\
{\bar g}_{ij}=\partial_{i}\phi^{\cal A}\partial_{j}\phi^{\cal B}G_{{\cal A}{\cal B}}(\phi)&=\bar e_{i}{}^{a}\bar e_{j}{}^{b}\delta_{ab}\ .\label{lsforphi}
\end{align}
The upper triangular vierbein \eqref{f:vierbeins} has 10 components, and is merely a repackaging of the 10 components of $\bar g_{\mu\nu}$(which in turn depend on the $\phi^{\cal A}$). 

For the physical metric $g_{\mu\nu}$, we parameterize its 16 component vierbein as a local Lorentz transformation (LLT) $\Lambda$, which has 6 components, times a vierbein $\hat E$ which is constrained in some way so that it has only 10 components,
\be\label{uptri} E_{\mu}{}^{A}=\Lambda^{A}{}_{B}\hat{E}_{\mu}{}^{B}\ee
The freedom to choose the constraints for $\hat E$ allows us to make different aspects of the theory manifest.  
The mixing term \eqref{mixaction}, in terms of vierbeins, takes the form
\begin{align}
S_{\rm mix}&\equiv -{M_{P}^{2}m^{2}\over 8}\sum_{n=1}^{3}{\beta_{n}\over n!(4-n)!}S^{(n)}_{\rm mix},\nonumber \\
S^{(1)}_{\rm mix}&=\int \epsilon_{ABCD} \bF^{A}\wedge\bE^{B}\wedge\bE^{C}\wedge\bE^{D},\nonumber \\
S^{(2)}_{\rm mix}&=\int \epsilon_{ABCD} \bF^{A}\wedge\bF^{B}\wedge\bE^{C}\wedge\bE^{D}, \nonumber\\
S^{(3)}_{\rm mix}&=\int \epsilon_{ABCD} \bF^{A}\wedge\bF^{B}\wedge\bF^{C}\wedge\bE^{D}.\label{mixterms}
\end{align}

The dynamical vierbein  
has 16 components, 6 more than the metric. 
If we choose the 6 constraints which $\hat E$ must satisfy to be the symmetry condition,
\be\label{symmcond}
\hat E_{\mu [A}\bar E^\mu_{\ B]}=0,
\ee
then we can show using the arguments in \cite{Hinterbichler:2012cn} (see also \cite{Deffayet:2012zc} for subtleties) that the extra 6 fields in $\Lambda$ are auxiliary fields which are eliminated by their own equations of motion, setting $\Lambda=1$, and the resulting theory is dynamically equivalent to the metric formulation \eqref{actionm}.

Instead, we take $\hat E$ to be of upper triangular form
\begin{align}
\hat{E}_{\mu}{}^{A}=\begin{pmatrix}
N& N^{i}e_{i}{}^{a}\\ 0 & e_{i}{}^{a}\end{pmatrix}. \label{e:vierbeins}
\end{align}
Here the spatial dreibein $e_i^{\ a}$ is arbitrary, containing 9 components.  The LLT  $\Lambda$ in \eqref{uptri} depends now on $3$ boost parameters $p^{a}$ and can be written as
\begin{align}
\Lambda(p)^A_{\ B} =\begin{pmatrix}
\gamma & p^{b}\\ p_{a}& \delta^{a}_{b}+\frac{1}{1+\gamma}p^{a}p_{b}
\end{pmatrix} \ ,
\label{uppertriangularLLT}
\end{align}
where $\gamma \equiv \sqrt{1+p^{a}p_{a}}$.  Using this decomposition, the 16 component vierbein $E_{\mu}{}^{A}$ is parameterized in terms of the 3 components of $p^{a}$, one $N$, 3 components of $N^{i}$ and the $9$ components of $e_{i}{}^{b}$.

{\bf Hamiltonian analysis:}
We start the Hamiltonian analysis by Legendre transforming with respect to the spatial vierbein $e_i^{\ a}$, defining canonical momenta $\pi^i_{\ a}={\partial {\cal L}\over\partial \dot e_i^{\ a}}$.
Since $S_{\rm mix}$ contains no time derivatives of the physical metric, and $S_{\rm Galileon}$ has no dependence on the physical metric at all, the expressions for the canonical momenta are the same as their GR counterparts.  In particular, there will be three primary constraints
\be
{\cal P}_{ab}=e_{i[a}\pi^i_{\ b]}=0 \,.
\label{rotconst}
\ee
In GR, these are first class constraints which generate local rotations.

The Einstein-Hilbert part of the action takes the form\footnote{See \cite{Deser:1976ay} or Appendix B of \cite{Hinterbichler:2012cn} for details of the Hamiltonian formulation of GR in vierbein form.}
\begin{align}
S_{\rm EH}
&=\int \rd^{4}x\, \pi^i_{\ a}\dot e_i^{\ a}-{1\over 2}\lambda^{ab}{\cal P}_{ab}- N\mathcal{C}\left (e,\pi\right ) -N^{j}\mathcal{C}_{j}\left (e,\pi\right ).
\end{align}
The anti-symmetric $\lambda^{ab}$ holds the three Lagrange multipliers for the three primary constraints \eqref{rotconst}.
The $N$ and $N^i$ appear as Lagrange multipliers enforcing respectively the Hamiltonian and momentum constraints of GR: $\mathcal{C}=0,\ \mathcal{C}_i=0$.  

We now look at the mixing terms \eqref{mixterms}.
The contributions to $\mathcal{L}_{\rm mix}$ are of the form
$\sim \epsilon^{\mu\nu\rho\sigma}\epsilon_{ABCD}E_{\mu}{}^{A}E_{\nu}{}^{B}{\bar E}_{\rho}{}^{C}{\bar E}_{\sigma}{}^{D}$, containing various numbers of copies of $E$ and $\bar E$.
From \eqref{f:vierbeins}, \eqref{e:vierbeins} and \eqref{uppertriangularLLT}, we see that the $\mu=0$ components of ${ E}_{\mu}{}^{A}$ and $\bar E_{\mu}{}^{A}$ are strictly linear in their respective lapses and shifts and the $\mu=i$ components are independent of the lapse and shift.  Therefore, due to the anti-symmetry of the epsilons, the entire mixing term is linear in the lapses and shifts, so we may write
\begin{align}
\mathcal{L}_{\rm mix}&= -N\mathcal{C}_{\rm mix}(e,\bar e,p)- N^{i}\mathcal{C}_{{\rm mix}, i}( e,\bar e,p)- \bar N\bar{\cal C}_{\rm mix}(e,\bar e,p)\nonumber \\
& \quad - \bar N^i\bar{\cal C}_{{\rm mix},i}( e,\bar e,p)-{\cal H}_{\rm mix}( e,\bar e,p)\ . 
\end{align}

The lapse and shift remain as Lagrange multipliers, and the $p^a$ appear algebraically.   We now solve the constraint enforced by $N^i$ for the $p^a$: $\mathcal{C}_{i}+\mathcal{C}_{{\rm mix}, i}=0\Rightarrow p^a=p^a(e,\bar e,\pi)$.  Plugging back into the action we obtain
\begin{align}
S
&=\int \rd^{4}x\, \pi^i_{\ a}\dot e_i^{\ a}-{1\over 2}\lambda^{ab}{\cal P}_{ab}- N\left[\mathcal{C}\left (e,\pi\right ) +\mathcal{C}_{\rm mix}(e,\bar e,\pi)\right] \nonumber\\ &- \bar N\bar{\cal C}_{\rm mix}(e,\bar e,\pi)  - \bar N^{i}\bar{\cal C}_{{\rm mix},i}(e,\bar e,\pi) \nonumber\\ &-{\cal H}_{\rm mix}(e,\bar e,\pi)+{\cal L}_{\rm Galileon}(\bar e,\bar N,\bar N^i)\ . \label{finallag}
\end{align}

It remains to Legendre transform with respect to the scalars $\phi^{\cal A}$, which appear through the dependence of $\bar N$, $\bar N^i$ and $\bar e_i^{\ a}$, as determined by \eqref{lsforphi}.  In order to avoid dealing with the complications of diffeomorphism invariance, we first fix unitary gauge, setting the first four fields equal to the space-time coordinates: $\phi^\mu=x^\mu$ (this can be done consistently in the action, since the missing equations of motion are implied by the remaining equations).  The Lagrangian \eqref{finallag} then depends on the remaining $(\dim\mathcal{M})-4$ scalars and their time derivatives.  Crucially, we see from \eqref{lsforphi} that while $\bar N$ and $\bar N_{i}$ depend on time derivatives of the scalars, the $\bar e_{i}{}^{a}$'s do not, and this in turn implies that the momenta conjugate to the scalars are independent of the dynamical lapse $N$.  Thus, when the scalar velocities are eliminated in terms of the momenta, the action will remain linear in $N$.  (If this were not the case, the lapse would no longer be a Lagrange multiplier, but would instead become an auxiliary field which does not impose a constraint on the remaining variables.)
The phase space is spanned by the 9 independent components of $e_i^{\ a}$, the physical scalars, and the canonical momenta.  Since the interaction terms break the local rotation invariance of GR, the 3 primary constraints \eqref{rotconst} associated with the local rotations will generate secondary constraints and form 3 second class pairs, thus removing 3 degrees of freedom.  The constraint enforced by $N$ is precisely the special primary constraint needed to remove the Boulware-Deser sixth degree of freedom, leaving 5 degrees of freedom for the massive graviton.  Analogously to what happens in massive gravity, we expect this special primary constraint to generate a secondary constraint to eliminate the ghost's conjugate momentum \cite{Hassan:2011ea}.

We have implicitly assumed that $S_{\rm Galileon}$ can be written in such a way that the $(\dim\mathcal{M})-4$ unitary gauge scalar fields appear with at most first time derivatives, so that we may define canonical momenta in the usual way.  This is not immediately obvious, because the higher order Galileons in $S_{\rm Galileon}$  possess higher derivative interactions.  However, the higher derivative interactions within $S_{\rm Galileon}$ are special in that they generate equations of motion which are no higher than second order in time.  This means it should be possible, after integrations by parts, to express these Lagrangians in terms of first time derivates only (though we shouldn't expect to be able to do the same with both the spatial and time derivates simultaneously). 
For example, take the case of a flat 5D target space, so that there is a single physical scalar $\phi$.  The unitary gauge induced metric is $\bar g_{\mu\nu}=\eta_{\mu\nu}+\partial_{\mu}\phi\,\partial_{\nu}\phi$.  The first higher-derivative Galileon is the cubic, coming from the extrinsic curvature term
\begin{align}
	S_K
		\sim \int {\rm d}^4 x \sqrt{-\bar g} \bar{K} \sim \int {\rm d}^4 x \frac{\partial_\mu\partial_\nu \phi\, \partial^\mu \phi \,\partial^\nu \phi}{1+(\partial \phi)^2} \ .
\end{align}
Looking at the structure of the possible higher-order time derivatives, the only offending term is
\begin{align}
	\frac{\ddot{\phi}\dot{\phi}^2}{1+ (\partial \phi)^2} \subset \mathcal{L}_K \ .
\end{align}
Expanding the denominator in powers of $(\partial \phi)^2$ we see that every term in this expansion is of the form $\ddot{\phi} \dot{\phi}^n (\vec{\nabla} \phi)^{2m}$ for some integer $m$ and $n$.  Integrating by parts, we can express each one in terms of first time derivatives only: $\ddot{\phi} \dot{\phi}^n (\vec{\nabla} \phi)^{2m}\sim\frac{{\rm d}}{{\rm d}t}(\dot{\phi}^{n+1})(\vec{\nabla} \phi)^{2m}\sim \dot{\phi}^{n+1}\frac{{\rm d}}{{\rm d}t}(\vec{\nabla} \phi)^{2m} $.  The same can be done with the higher Galileons and with a curved target space (see for example the Hamiltonian analysis of \cite{Zhou:2010di,Sivanesan:2011kw} in the non-relativistic case).

{\bf Conclusions:}
There exists~\cite{Goon:2011uw,Goon:2011qf,Goon:2011xf} a wide range of novel scalar field theories with interesting properties such as Vainshtein screening and non-renormalization theorems in common with the original Galileon models of ~\cite{Luty:2003vm,Nicolis:2008in}. These properties hold out the hope that such models may be of use both in particle physics, and as a possible way to modify gravity in the infrared. However, coupling such fields to General Relativity in a way that preserves their symmetries and second order equations of motion seems to be impossible~\cite{Deffayet:2009wt}. Instead, Galileon-like scalar fields seem to most naturally couple to dRGT massive gravity~\cite{Gabadadze:2012tr}.

The consistency of such a proposal rests on the preservation of the hard-won ghost-free structure of the dRGT theory. 
In this letter, we have shown for the first time that a theory of nonlinear massive gravity coupled to DBI scalars in such a way as to preserve the generalized Galileon shift symmetry and the property of having second order equations of motion is ghost free. Our proof is based on the vierbein formulation of massive gravity, in which the Hamiltonian analysis simplifies. Our analysis shows that the dRGT-DBI system provides a consistent framework in which models of interest to cosmology~\cite{Hinterbichler:2013dv} may be developed.

{\bf Acknowledgments:}
Work at the University of Pennsylvania was supported in part by the US Department of Energy, and NASA ATP grant NNX11AI95G. Research at Perimeter Institute is supported by the Government of Canada through Industry Canada and by the Province of Ontario through the Ministry of Economic Development and Innovation. The work of KH was made possible in part through the support of a grant from the John Templeton Foundation. The opinions expressed in this publication are those of the authors and do not necessarily reflect the views of the John Templeton Foundation.


\begin{thebibliography}{99}

\bibitem{Fierz:1939ix}
  M.~Fierz and W.~Pauli,
  Proc.\ Roy.\ Soc.\ Lond.\  A {\bf 173}, 211 (1939).


\bibitem{Boulware:1973my} 
  D.~G.~Boulware and S.~Deser,
  Phys.\ Rev.\ D {\bf 6}, 3368 (1972).


\bibitem{deRham:2010ik} 
  C.~de Rham and G.~Gabadadze,
  Phys.\ Rev.\ D {\bf 82}, 044020 (2010)
  [arXiv:1007.0443 [hep-th]].


\bibitem{deRham:2010kj} 
  C.~de Rham, G.~Gabadadze and A.~J.~Tolley,
  Phys.\ Rev.\ Lett.\  {\bf 106}, 231101 (2011)
  [arXiv:1011.1232 [hep-th]].


\bibitem{Hassan:2011hr} 
  S.~F.~Hassan and R.~A.~Rosen,
  Phys.\ Rev.\ Lett.\  {\bf 108}, 041101 (2012)
  [arXiv:1106.3344 [hep-th]].


\bibitem{Hassan:2011ea} 
  S.~F.~Hassan and R.~A.~Rosen,
  JHEP {\bf 1204}, 123 (2012)
  [arXiv:1111.2070 [hep-th]].


\bibitem{deRham:2011qq} 
  C.~de Rham, G.~Gabadadze and A.~J.~Tolley,
  JHEP {\bf 1111}, 093 (2011)
  [arXiv:1108.4521 [hep-th]].


\bibitem{deRham:2011rn} 
  C.~de Rham, G.~Gabadadze and A.~J.~Tolley,
  Phys.\ Lett.\ B {\bf 711}, 190 (2012)
  [arXiv:1107.3820 [hep-th]].


\bibitem{Mirbabayi:2011aa} 
  M.~Mirbabayi,
  Phys.\ Rev.\ D {\bf 86}, 084006 (2012)
  [arXiv:1112.1435 [hep-th]].


\bibitem{Hinterbichler:2012cn} 
  K.~Hinterbichler and R.~A.~Rosen,
  JHEP {\bf 1207}, 047 (2012)
  [arXiv:1203.5783 [hep-th]].


\bibitem{ArkaniHamed:2002sp} 
  N.~Arkani-Hamed, H.~Georgi and M.~D.~Schwartz,
  Annals Phys.\  {\bf 305}, 96 (2003)
  [hep-th/0210184].


\bibitem{Gabadadze:2012tr} 
  G.~Gabadadze, K.~Hinterbichler, J.~Khoury, D.~Pirtskhalava and M.~Trodden,
  Phys.\ Rev.\ D {\bf 86}, 124004 (2012)
  [arXiv:1208.5773 [hep-th]].


\bibitem{deRham:2010eu} 
  C.~de Rham and A.~J.~Tolley,
  JCAP {\bf 1005}, 015 (2010)
  [arXiv:1003.5917 [hep-th]].


\bibitem{Hinterbichler:2010xn} 
  K.~Hinterbichler, M.~Trodden and D.~Wesley,
  Phys.\ Rev.\ D {\bf 82}, 124018 (2010)
  [arXiv:1008.1305 [hep-th]].


\bibitem{Goon:2011qf} 
  G.~Goon, K.~Hinterbichler and M.~Trodden,
  JCAP {\bf 1107}, 017 (2011)
  [arXiv:1103.5745 [hep-th]].


\bibitem{Goon:2011uw} 
  G.~Goon, K.~Hinterbichler and M.~Trodden,
  Phys.\ Rev.\ Lett.\  {\bf 106}, 231102 (2011)
  [arXiv:1103.6029 [hep-th]].


\bibitem{Burrage:2011bt} 
  C.~Burrage, C.~de Rham and L.~Heisenberg,
  JCAP {\bf 1105}, 025 (2011)
  [arXiv:1104.0155 [hep-th]].

\bibitem{toappear} 
  G.~Goon, K.~Hinterbichler and M.~Trodden,
 to appear


\bibitem{Deffayet:2009wt} 
  C.~Deffayet, G.~Esposito-Farese and A.~Vikman,
  Phys.\ Rev.\ D {\bf 79}, 084003 (2009)
  [arXiv:0901.1314 [hep-th]].


\bibitem{Deffayet:2009mn} 
  C.~Deffayet, S.~Deser and G.~Esposito-Farese,
  Phys.\ Rev.\ D {\bf 80}, 064015 (2009)
  [arXiv:0906.1967 [gr-qc]].


\bibitem{Hassan:2011zd} 
  S.~F.~Hassan and R.~A.~Rosen,
  JHEP {\bf 1202}, 126 (2012)
  [arXiv:1109.3515 [hep-th]].
  
  
\bibitem{Deffayet:2012nr} 
  C.~Deffayet, J.~Mourad and G.~Zahariade,
  JCAP {\bf 1301}, 032 (2013)
  [arXiv:1207.6338 [hep-th]].

\bibitem{Deser:2012qx} 
  S.~Deser and A.~Waldron,
  arXiv:1212.5835 [hep-th].

\bibitem{Deffayet:2012zc} 
  C.~Deffayet, J.~Mourad and G.~Zahariade,
  arXiv:1208.4493 [gr-qc].


\bibitem{Deser:1976ay} 
  S.~Deser and C.~J.~Isham,
  Phys.\ Rev.\ D {\bf 14}, 2505 (1976).


\bibitem{Zhou:2010di} 
  S.~-Y.~Zhou,
  Phys.\ Rev.\ D {\bf 83}, 064005 (2011)
  [arXiv:1011.0863 [hep-th]].


\bibitem{Sivanesan:2011kw} 
  V.~Sivanesan,
  Phys.\ Rev.\ D {\bf 85}, 084018 (2012)
  [arXiv:1111.3558 [hep-th]].


\bibitem{Goon:2011xf} 
  G.~Goon, K.~Hinterbichler and M.~Trodden,
  JCAP {\bf 1112}, 004 (2011)
  [arXiv:1109.3450 [hep-th]].


\bibitem{Luty:2003vm} 
  M.~A.~Luty, M.~Porrati and R.~Rattazzi,
  JHEP {\bf 0309}, 029 (2003)
  [hep-th/0303116].


\bibitem{Nicolis:2008in} 
  A.~Nicolis, R.~Rattazzi and E.~Trincherini,
  Phys.\ Rev.\ D {\bf 79}, 064036 (2009)
  [arXiv:0811.2197 [hep-th]].


\bibitem{Hinterbichler:2013dv} 
  K.~Hinterbichler, J.~Stokes and M.~Trodden,
  arXiv:1301.4993 [astro-ph.CO].
\end{thebibliography}
\end{document}